\documentclass{revtex4}

\usepackage{amsmath}
\usepackage{amssymb}

\begin{document}

\title{Soft singularity and the fundamental length}
\author{Vladimir Dzhunushaliev
\footnote{Senior Associate of the Abdus Salam ICTP}
} 
\email{dzhun@hotmail.kg} \affiliation{
Dept. Phys. and Microel. Engineer., Kyrgyz-Russian Slavic
University, Bishkek, Kievskaya Str. 44, 720021, Kyrgyz Republic\\
}
\author{Ratbay Myrzakulov }
\email{cnlpmyra@satsun.sci.kz} \affiliation{Institute of Physics and 
Technology, 480082, Almaty-82, Kazakhstan}


\begin{abstract}
It is shown that some regular solutions in 5D Kaluza-Klein gravity 
may have interesting properties if one from the parameters is in the 
Planck region. In this case the Kretschman metric invariant runs up 
to a maximal reachable value in nature, i.e. practically the metric 
becomes singular. This observation allows us to suppose that in this 
situation the problems with such soft singularity will be much 
easier resolved in the future quantum gravity then by the situation 
with the ordinary hard singularity (Reissner-Nordstr\"om 
singularity, for example). It is supposed that the analogous 
consideration can be applied for the avoiding the hard singularities 
connected with the gauge charges. 
\end{abstract}

\maketitle

\section{Introduction}

Any solution in any theory has some parameters: mass, charge, 
angular momentum, characteristic length and so on. The properties of 
the solution depends, of course, on the value of the parameters: the 
presence of event horizon, singularity and so on. In this paper we 
investigate the case when these properties crucially depend on the 
parameters values. The reason for this is that the nature may have a 
natural length: Planck (fundamental) length. In this case one can 
expect that some properties of the solution will be changed on the 
level of the fundamental length. 
\par 
Here we consider the solution of the 5D Kaluza-Klein gravity 
\cite{Chodos:1980df} - \cite{Clement:1986dn} which has two 
parameters: electric charge $q$ and some characteristic length 
$r_0$. From the mathematical point of view this solution is regular 
everywhere. We will show that if $r_0 \approx \ell$ ($\ell \approx 
10^{-33}cm$ is the Planck length) then from the physical point of 
view the situation completely changes: in a part of this spacetime 
the Kretschman invariant becomes $\approx 1/\ell^4$ and this means 
that it is a singularity since this value is maximal if the 
fundamental length does exist. 

\section{Fundamental length}

In 50-th Wheeler has introduced the notion of a Planck length as a 
minimal achievable length in nature. The physical consequences of 
the existence of this length are very important. For example, the 
fluctuations of the metric on this level probably lead to the 
appearance of a spacetime foam where the fluctuations of a spacetime 
topology takes place. Unfortunately the mathematical rigorous 
introduction of this length is possible in a quantum gravity theory 
only. Nevertheless using the theory of deformations of a Lie algebra 
connected with given physical theory one can show 
\cite{VilelaMendes:1994zg}-\cite{Faddeev} that a fundamental length 
is an essential ingredient of an extension of a physical theory. 
\par 
The essence of the deformation theory is that one can extent the 
commutator $[\; , \;]_0$ of a Lie algebra $\mathcal L_0$ by 
introducing a new parameter $t$ 
\begin{equation}
    \left[ A, B \right]_t = \left[ A, B \right]_0 +
    \sum^\infty_{i=1} M_i \left( A, B \right) t^i
\label{sec0-10}
\end{equation}
where $A,B \in V$ and $M_i \left( A, B \right) \in V$; $[\; , \;]_t$ 
is a new commutator corresponding to the new parameter $t$ (it can 
be, for example, either the speed of light $c$ or the Planck 
constant $\hbar$ or another completely new fundamental constant) in 
a new Lie algebra $\mathcal L_t$.
\par
A deformation of $\mathcal L_0$ is trivial if the algebra $\mathcal 
L_t$ is isomorphic to $\mathcal L_0$, i.e. there is an invertible 
transformation $T_t : V \rightarrow V$ such that
\begin{equation}
    T_t\left(
        \left[ A, B \right]_t
    \right) =
    \left[
        T_tA, T_tB
    \right]_0 .
\label{sec5-50}
\end{equation}
The most interesting case is with the non-isomorphic deformation 
from $\mathcal L_0$ to $\mathcal L_t$. In this case we can obtain a 
new physical theory with a new fundamental constant $t$. 
\par 
Moving by such a way one can show that: (a) the stabilization from 
the classical mechanics to the relativistic one is possible by 
introducing a new fundamental constant - the speed of light $c$; (b) 
the stabilization from the classical mechanics to the quantum one is 
possible with the introduction of a new fundamental constant - the 
Planck constant $\hbar$; (c) stabilization from classical 
relativistic mechanics to an algebra with a fundamental length is 
possible \cite{VilelaMendes:1994zg}-\cite{Chryssomalakos:2004gk}. 
\par 
In Ref's \cite{Ahluwalia-Khalilova:2005km}, 
\cite{Ahluwalia-Khalilova:2005jn} (resting on the principles of the 
deformation idea) it is shown that the Lie algebra for the interface 
of the gravitational and quantum realms is the stabilized form of 
the Poincar\'{e}-Heisenberg algebra which carries three additional 
parameters and one from them is the fundamental length. It is shown 
that the stable Snyder-Yang-Mendes Lie algebra is a serious 
candidate for the symmetries underlying freely falling frames at the 
interface of quantum and gravitational ideas. 

\section{The metric}

We start with the 5D metric 
\begin{equation}
    ds^2 = \frac{1}{\Delta(r)} dt^{2} - r_0^2 \Delta(r) \left [d\chi +  
    \omega (r)dt \right ]^2  - dr^{2} - 
    a(r)(d\theta ^{2} + \sin ^{2}\theta  d\varphi ^2).
\label{sec1-10}
\end{equation}
The 5D Einstein's equations are 
\begin{equation} 
    R_{AB} - \frac{1}{2}G_{AB} R = 0 
\label{sec1-45} 
\end{equation}
where $A,B = 0,1,2,3,5$. The solution which was found in Ref's 
\cite{Chodos:1980df} - \cite{Clement:1986dn} we present in the form 
\cite{dzhsin1}
\begin{eqnarray} 
    a & = & r^{2}_{0} + r^{2},
\label{sec1-20}\\
        \Delta & = & \frac{q}{2r_0} \frac{r^2 - r^2_0}{r^2 + r^2_0},
\label{sec1-30}\\
    \omega & = & \frac{4r_0}{q} \frac{r}{r^2_0 - r^2} .
\label{sec1-40}
\end{eqnarray}
The curvature scalar $R$ and the invariant $R_{AB} R^{AB}$ are zero 
in the consequence of the 5D Einstein's equations \eqref{sec1-45}. 
The corresponding Kretschman invariant for the metric 
\eqref{sec1-10} and the solution \eqref{sec1-20}-\eqref{sec1-40} is 
\begin{equation}
    K = R^{ABCD} R_{ABCD} = 
    \frac{24}{r_0^4} 
    \frac{\displaystyle{1 - 2\left( \frac{r}{r_0} \right)^2}}
    {\displaystyle{\left[ 
        1 + \left( \frac{r}{r_0} \right)^2 
    \right]^4}} .
\label{sec1-50} 
\end{equation}
We see that the spacetime with the metric 
\eqref{sec1-20}-\eqref{sec1-40} is regular everywhere. But there is 
one subtlety. This statement is correct only if nature has not any 
fundamental length. If such length exists then we should be more 
careful. The parameter $r_0$ describes the region of the spacetime 
\eqref{sec1-10}-\eqref{sec1-40} between two surfaces where $ds^2 
\left( \pm r_{0} \right) = 0$. In contrast with the Schwarzschild 
metric 
\begin{equation}
    ds^2 = \left(
        1 - \frac{r_g}{r}
    \right) dt^2 - 
    \frac{dr^2}{\displaystyle{1 - \frac{r_g}{r}}} - 
    r^2 \left( d \theta^2 + \sin^2 \theta d \phi^2 \right)
\label{sec1-60} 
\end{equation}
we do not need to introduce a new coordinate system to describe the 
region with $|r| < r_0$. It is visible that the metric 
\eqref{sec1-10}, \eqref{sec1-20}-\eqref{sec1-40} can be written in 
the form 
\begin{equation}
    ds^2 = \frac{2 r_0}{q} \frac{r^2 - r_0^2}{r^2 + r_0^2} 
    \left( dt^2 - \frac{q^2}{4} d \chi^2 \right) + 
    \frac{r_0^2 r}{r^2 + r_0^2} dt d\chi - 
    dr^2 - 
    \left( r^2 + r_0^2 \right) 
    \left( d \theta^2 + \sin^2 \theta d \phi^2 \right) .
\label{sec1-70} 
\end{equation}
Near to the surface $r = \pm r_0$ 
\begin{equation}
    ds^2 = \frac{r_0}{2} dt d\chi - dr^2 - 
    2 r_0^2 \left( d \theta^2 + \sin^2 \theta d \phi^2 \right) + 
    \mathcal{O} 
    \bigl( r - r_0 \bigl) 
\label{sec1-80} 
\end{equation}
here immediately we see that we have not any singularity at the 
points $r = \pm r_0$.

\section{Soft singularity} 

The essential thing here is that if $r_0 = \ell$ then the Kretschman 
invariant \eqref{sec1-50} blow up in the region 
$\left| r \right| \leq r_0$ 
\begin{equation}
    K \approx \frac{1}{\ell^4} .
\label{sec2-10} 
\end{equation}
As the Planck length $\ell$ is the minimal length then the maximal 
possible value of $K$ can be $1 / \ell^4$ only. This statement leads 
to the very strong conclusion: \emph{an external observer sees the 
region $\left| r \right| \leq r_0$ as a soft singularity.} The word 
``soft'' means that it is not ordinary (hard) singularity where $K = 
\infty$. But here in the presence of the minimal length we can say 
that $\frac{1}{\ell^4} \approx \infty$. 
\par 
Now we will try to understand how one can apply this result to 
physics.

\subsection{Inner structure of a singularity}

At first we would like to describe what we have for this situation 
with $r_0 \approx \ell$. An external observer sees that the volume 
$r \leq r_0$ has a singularity although from the mathematical point 
of view the metric is regular one. The resolution of this problem is 
that we consider the \emph{classical} solutions but one can hope 
that inside of the region $|r| \leq r_0$ a quantum gravity can gives 
us some nonsingular answer. The main hope here is that even on the 
classical level we have some inner structure of the soft singularity 
consequently the quantum gravity must gives us some nonsingular 
answer on the problem of an inner structure of the singularity. 
\par 
In fact the above-mentioned consideration gives us a hint that any 
singularity where there is a gauge charge (electric or color) 
actually has an inner nonsingular structure. The regularization 
process of such singularity can be connected with the fact that near 
to the (even strong) singularity the gravitational field \emph{in 
the presence of a gauge field} becomes so strong that it excites the 
dynamics on the extra dimensions and this dynamics can be described 
on the language of quantum gravity only. 

\subsection{Very naive model of the electric charge}

The metric \eqref{sec1-10}, \eqref{sec1-20}-\eqref{sec1-40} can be 
considered as a 5D model of the electric charge since far away from 
the origin $(r = 0)$ the electric field becomes like Coulomb 
\begin{equation}
    E = \omega' \approx \frac{1}{r^2}  , \; 
    \left| r \right| \gg r_0 .
\label{sec3-10} 
\end{equation}
The problem here is that there is a modification of the Coulomb law 
which can be measured experimentally. One way for the reduction of 
this correction to the minimum is $r_0 \rightarrow 0$. On this way 
we have only one obstacle: the fundamental length, i.e. $\left( r_0 
\right )_{min} \approx \ell$. In this case we will have the soft 
singularity and all that we spoke earlier in this occasion. Let us 
to consider the case with $r_0 = \ell$. At the first we would like 
to analyze carefully the notion of the electric field in this case. 
\par 
The 5D $(\chi t)$-Einstein's equation (4D Maxwell equation) is taken 
as having the following solution
\begin{equation}
  \omega ' = \frac{q}{a \Delta ^2} .
\label{sec3-20}
\end{equation}
For the determination of the physical sense of the constant $q$ let 
us write the $(\chi t)$-Einstein's equation in the following way :
\begin{equation}
    \left( \omega ' \Delta ^2 4 \pi a \right)' = 0.
\label{sec3-30}
\end{equation}
The 5D Kaluza - Klein gravity after the dimensional reduction 
indicates that the Maxwell tensor is
\begin{equation}
    F_{\mu \nu} = \partial_\mu A_\nu - \partial _\nu A_\mu .
\label{sec3-40}
\end{equation}
That allows us to write the electric field as $E_r = \omega '$. 
Eq.\eqref{sec3-30}, with the electric field defined by 
\eqref{sec3-40}, can be compared with the Maxwell's equations in a 
continuous medium
\begin{equation}
    \mathrm{div} \mathbf {D} = 0
\label{sec3-50}
\end{equation}
where $\mathbf {D} = \epsilon \mathbf {E}$ is an electric 
displacement and $\epsilon$ is a dielectric permeability. Comparing 
Eq. \eqref{sec3-50} with Eq. \eqref{sec3-30} we see that the 
magnitude $q/a = \omega ' \Delta^2 $ is like to the electric 
displacement and the dielectric permeability is $\epsilon = \Delta^2 
$. This means that $q$ can be taken as the Kaluza-Klein electric 
charge because the flux of the electric displacement is $\Phi = 4\pi 
a \mathrm D = 4\pi q$.
\par
For the 4D observer at the infinity $(r \rightarrow \infty)$ we have 
the following asymptotical behavior of the electric displacement 
$\mathrm D_r$ and the scalar field $G_{55} = \Delta$ 
\begin{eqnarray}
  \mathrm D_r &=& \frac{q}{r^2 + r_0^2} \approx \frac{q}{r}
  \left( 1 - \frac{\ell^2}{r^2} \right) , 
\label{sec3-60}\\
  \Delta &=& \frac{q}{2r_0} \frac{r^2 - r_0^2}{r^2 + r_0^2} 
  \approx \frac{q}{2\ell} 
  \left( 1 - 2 \frac{\ell^2}{r^2} \right)
\label{sec3-70}
\end{eqnarray}
as $r_0 \approx \ell$. The deviation \eqref{sec3-60} from the 
Coulomb law can not be measured with the modern technical abilities. 
For the interpretation of the 5D Kaluza-Klein gravity as 4D 
electrogravity it is very important to have the dilaton field 
$G_{55}=1$. From the equation \eqref{sec3-70} we see that it is true 
with the accuracy $\approx \ell^2 / r^2$. Probably the most 
important here is that the 4D observer will see the (soft) 
singularity at the center. 
\par 
It is interesting to compare this situation with the 4D 
Reissner-Nordst\"om solution. In the last case the metric crucially 
depends on the relation between the mass $m$ and the electric charge 
$q$. If $m^2 > q^2$ ($c = G = 1$) then there is an event horizon and 
in opposite case a naked singularity. In our 5D case the 4D metric 
is 
\begin{equation}
    ds^2 = \frac{1}{\Delta(r)} dt^{2} - dr^{2} - 
    a(r)(d\theta ^{2} + \sin ^{2}\theta  d\varphi ^2)
\label{sec3-80}
\end{equation}
and we see that $G_{tt}$ metric component \eqref{sec1-70} 
asymptotically is 
\begin{equation}
    g_{tt} \approx \frac{2 \ell}{q} 
    \left(
        1 + \frac{2 \ell^2}{r^2}
    \right).
\label{sec3-90}
\end{equation}
One can redefine the time $t$ to remove the factor $2\ell /q$. From 
this equation we see that we have not the term $m/r$ that means that 
the mass is zero ($m = 0$) for this 4D interpretation of the 5D 
metric \eqref{sec1-10}. 
\par 
Near to the region $|r| \leq r_0$ we have the soft singularity which 
hopefully will be regularized in a future quantum gravity theory. 
After that we will have regular everywhere spacetime with the finite 
electric field, finite energy of electric field, arbitrary value of 
electric charge $q$ in contrast with the Reissner-Nordstr\"om 
solution where the topological structure of spacetime depends on the 
relation between mass $m$ and charge $q$. 

\section{Conclusions}

In this notice we have shown that the well known spherically 
symmetric solution in the 5D Kaluza-Klein gravity has unexpected 
properties if one of the parameters of this solution is in the 
Planck region. This conclusion essentially follows from the fact 
that a minimal length exists in nature. We have shown that in this 
case an external observer will see the soft singularity. The 
appearance of such kind singularity is connected with the fact that 
at the center of this spacetime the Kretschman metric invariant blow 
up from zero (at $r = r_0 / \sqrt 2$) up to infinity (at $r = 0$). 
Certainly such situation demand the careful consideration on the 
basis of quantum gravity. The good news here is that such (soft) 
singularity has an inner structure in contrast with the hard 
singularity (Reissner-Nordstr\"om solution, for example). It gives a 
hope that the resolution of the singularity problem in general 
relativity can be found in quantum gravity. 
\par 
Mathematically the soft singularity is regular one since near to 
$r=r_0$ the gravity is so strong that the dynamic on the 5-th 
dimension is excited. This observation permits us to suppose that 
\emph{any} singularity having gauge charge (electric or color) can 
be regularized by the similar way: near to the singularity the 
superstrong gravitational field excites the metric dynamic on the 
extra dimensions and the quantum gravity effects smoothes away the 
singular metric up to a regular quantum one.

\end{document}